\documentclass{PoS}

\title{Observations of IntraDay Variable sources with the Effelsberg and Urumqi Radio Telescopes}

\ShortTitle{Observations of IDV sources}

\author{\speaker{N. Marchili}, T. P. Krichbaum\\
        Max-Planck-Institut f\"ur  Radioastronomie, Bonn, Germany\\
        E-mail: \email{marchili@mpifr-bonn.mpg.de}, \email{tkrichbaum@mpifr-bonn.mpg.de}}

\author{X.Liu, H. G. Song\\
        Urumqi Observatory, National Astronomical Observatories, Chinese Academy
        of Sciences, Urumqi 830011, PR China\\
        E-mail: \email{liux@uao.ac.cn}, \email{songhg@uao.ac.cn}}

\author{K. \'E. Gab\'anyi\\
        HAS, Research Group for Geodesy and Geodynamics, Budapest, Hungary\\
        F\"OMI, Satellite Geodetic Observatory, Penc, Hungary\\
        E-mail: \email{gabanyik@sgo.fomi.hu}}

\author{L. Fuhrmann, P. M\"uller, A. Witzel, J. A. Zensus\\
        Max-Planck-Institut f\"ur  Radioastronomie, Bonn, Germany\\
        E-mail: \email{fuhrmann@mpifr-bonn.mpg.de},
        \email{peter@mpifr-bonn.mpg.de}, \email{awitzel@mpifr-bonn.mpg.de}, 
        \email{azensus@mpifr-bonn.mpg.de}} 

\author{J. L. Han\\
        National Astronomical Observatories, Chinese Academy of Sciences, 
        Beijing 100012, PR China\\
        E-mail: \email{hjl@bao.ac.cn}}

\abstract{A sample of classical IntraDay Variable (IDV) and IDV candidate sources has been 
monitored with the Urumqi 25m telescope and the Effelsberg 100m telescope. Aim of the project is to investigate the origin of IntraDay Variability, a phenomenon which has been observed in about 30\% of flat spectrum radio quasars. Simultaneous Effelsberg-Urumqi observations demonstrated that the Urumqi antenna, although relatively small in diameter, is well suitable for IDV experiments. A few Urumqi datasets, however, turned out to be affected by a spurious $\sim 24$ hours modulation, an effect which has been removed by means of a new procedure for data reduction. In about 14 months, 12 epochs of observation have been collected, for a total observing time of more than 45 days. The epochs are regularly distributed over the whole year, in order to check for the presence of systematic annual changes in the variability time scales - a crucial test for verifying the consistency of source-extrinsic models of the variability. Preliminary time-analysis of the monitored sources revealed some hint for a slowing down of the characteristic time scales of S5~0716+714, a result that, if confirmed, would be compatible with a source-extrinsic origin of the variability, in contrast to previous IDV studies. No significant modulation of the time scales has been detected for S4~0954+658.}

\FullConference{Bursts, Pulses and Flickering:Wide-field monitoring of the dynamic radio sky\\
		 June 12-15 2007\\
		 Kerastari, Tripolis, Greece}

\begin{document}

\section{Introduction}

The discovery of IntraDay Variability (IDV) in the radio flux from AGNs (\cite{Witzel 1986}) gave rise to a series of fundamental issues concerning the origin of the variability. The observation of flux variations which for some sources can be as high as 30\% of the mean flux, on time scales of the order of a few hours, led to the question whether such extreme phenomena can physically be intrinsic to sources. From causality arguments, the variability time scales constrain the maximum size of the emitting region. For IDV sources, this frequently implies very high brightness temperatures, far exceeding the Inverse Compton limit (see \cite{Pauliny}, \cite{Crusius}). In a source-intrinsic interpretation, only very high Doppler boost factors - of the order of a few tens, or even hundreds - or particular geometries of the emitting regions can explain such short time scale variability (\cite{Qian}, \cite{Qian2}). No strong evidence in favour of the second hypothesis has been found yet, while the first one seems to contradict the results of VLBI observations (\cite{Kellermann}).

More than twenty years after the first detection of IDV,
% - e.g. through intensive flux monitoring of IDV sources or the direct 
% observation through VLBI techniques - 
different models have been developed for the explanation of the variability, but the picture concerning the nature of IDV is still not homogeneous. In its most extreme manifestation - the fast scintillators, e.g. J1819+3845 and PKS~0405$-$385, varying on time scales of less than a few hours - the variability is most likely due to InterStellar Scintillation (ISS), an extrinsic effect whose clearest signature is an {\it annual modulation of the variability time scales}, due to the variation of the relative velocity between the Earth and the projected velocity of the scattering screen (\cite{Dennett1}, \cite{Dennett2}, \cite{Bignall}). Some evidence of annual modulation was also found in the quasar 0917+624 (\cite{Rickett}) - a type II-IDV source (\cite{Quirrenbach2}), characterized by slower time scales than fast scintillators - but still problems remain in the interpretation of its polarization variations (\cite{Qian3}). In the case of S5~0716+714, another type II source, simultaneous optical-radio IDV observations revealed a tight correlation between the two bands (see \cite{Wagner}, \cite{Quirrenbach 1991}), a result which has been interpreted as evidence for source-intrinsic variability.

It is likely that both extrinsic and intrinsic effects play a role in the phenomenon, but the fundamental issues about the origin of IDV and the intrinsic source size are still without solution.

\subsection{The project}
  
In December 2004, a project for the intensive monitoring of IDV sources was started at the Urumqi radio telescope (China). The Urumqi telescope is a 25-meter antenna, equipped with a 4.8GHz state-of-the-art receiver, provided, as well as the new telescope driving program, by the MPIfR. Aim of the project, which is still on-going, is a systematic study of the {\it changes} - if any - in the variability pattern of selected IDV sources throughout the year. The detection and proper estimation of regular variations could be used for discriminating between existing variability models, and for constraining their parameters.

  \begin{figure}[ht]
   \centering
   \includegraphics[width=8cm]{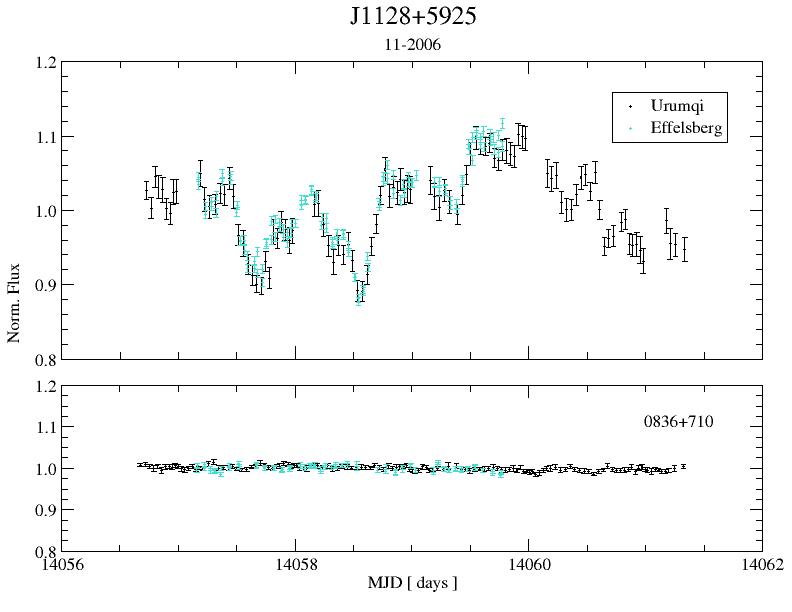}
      \caption{The variability curves of J1128+5925 (top panel) and a secondary calibrator (0836+710, bottom panel) from simultaneous observations at Effelsberg and Urumqi. The agreement between the curves is very good, also when the source shows fast variability.}
         \label{1128-11.06}
   \end{figure}

 The key-points of the project are the regularity and the short separation between consecutive observations, and the duration of each experiment, which has to be long enough to ensure a proper estimation of the time scales.

 Main targets of the observations are some archetypical type II-IDV sources such as S5~0716+714, S4~0917+624 and S4~0954+658, and the recently discovered source J1128+5925, whose variability characteristics seem to bridge type II and fast scintillators (see \cite{Gabanyi}). Some new IDV candidates have been sporadically included in the observed sample as well. 

Several simultaneous Effelsberg-Urumqi observations have been carried out during the year 2006, for cross-checking the results from two telescopes with different properties. All the observations have been performed at 6cm wavelength - using the 4.85GHz receiver in Effelsberg and the 4.8GHz receiver in Urumqi, with bandwidths of respectively 500MHz and 320MHz.  The degree of correlation between the data provided by the two antennas is extremely satisfying (see Fig. \ref{1128-11.06}), showing that the smaller diameter of the Urumqi telescope does not represent a severe limit for performing sensitive IDV experiments.

%  
%__________________________________________________________________

\section{The Observations}

So far, 12 epochs of observation have been collected and analyzed, summing up to a total observing time of more than 45 days. In each epoch, IDV sources have been observed along with two classes of calibrators - primary and secondary. The former includes objects showing no flux variability at all, and are used for estimating the conversion factor from K to Jy; the latter includes objects close to IDV sources, showing no flux variability on time scales of days; they are used for removing spurious effects due to residual receiver gain variations and changes in the weather.

 Relevant information about the observations at each epoch are summarized in Table \ref{epochs}. The average time sampling, in column 3, denotes the number of data points per hour, independently from the observed source. The duty cycle for IDV sources is reported in column 5; it is worth noting that the duty cycle for IDV sources was generally higher than for calibrators.

\begin{table}[t]
\centering
%\vspace{3cm}
\begin{tabular}{|l|c|c|c|c|c|}
\hline
        &          &                   &                   & Duty cycle \\
 Epoch  & Duration & Average sampling  & Number of sources & for IDV sources \\
        &    (d)   &$data\cdot(h^{-1}$)&                   & $data\cdot(h^{-1}$)\\
\hline
15.03-18.03.2006   &    3.0   &      10.0         &         18        & 0.7 \\
27.04-01.05.2006   &    3.9   &      ~8.7         &         14        & 0.7 \\
09.06-12.06.2006   &    3.2   &      10.9         &         11        & 1.2 \\
14.07-18.07.2006   &    4.0   &      ~8.6         &         11        & 0.9 \\
19.08-25.08.2006   &    6.4   &      11.3         &         12        & 1.3 \\
23.09-28.09.2006   &    5.0   &      11.0         &         12        & 1.2 \\
17.11-22.11.2006   &    4.7   &      12.0         &         14        & 1.2 \\
18.12-21.12.2006   &    2.4   &      11.0         &         12        & 1.3 \\
25.01-27.01.2007   &    2.3   &      11.2         &         14        & 1.1 \\
12.02-16.02.2007   &    4.0   &      11.1         &         15        & 1.0 \\
24.03-27.03.2007   &    2.8   &      11.0         &         16        & 0.9 \\
20.04-24.04.2007   &    3.7   &      ~7.9         &         16        & 0.8 \\
\hline
\end{tabular}
\caption{12 epochs of observations with the Urumqi telescope: in column 1 are reported the starting and ending dates of the experiments; in column 2, the duration; in column 3, the mean number of flux density measurements per hour; in column 4, the number of observed sources; in column 5, the average duty cycle for IDV sources.}
\label{epochs}
\end{table}

\subsection{Data reduction}

The reduction of the Urumqi data follows the same methods successfully applied for the IDV measurements performed with the Effelsberg telescope (for a detailed description, see \cite{Kraus} and \cite{Gabanyi}):

\begin{itemize}
    \item flux density measurements by gaussfits to cross-scans;
    \item correction for the opacity of the atmosphere (when needed);
	 \item correction for the average pointing offset of each scan;
	 \item correction for the gain-elevation effect;
    \item correction for the gain-time effect (due, e.g., to changes in the weather or in system temperature);
    \item conversion of the flux density from K to Jy.
\end{itemize}

  \begin{figure}[!ht]
   \centering
   \includegraphics[width=6.5cm]{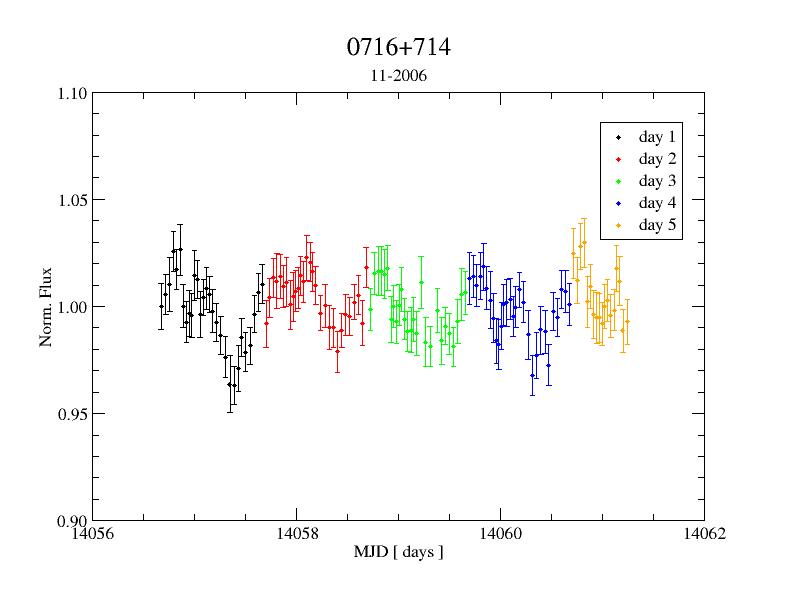}
\hspace{1cm}
   \includegraphics[width=6.5cm]{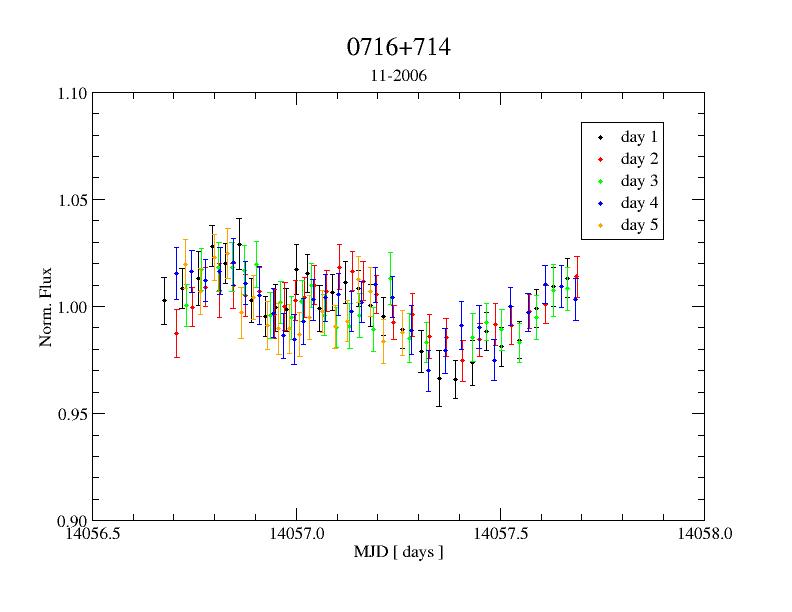}

   \includegraphics[width=6.5cm]{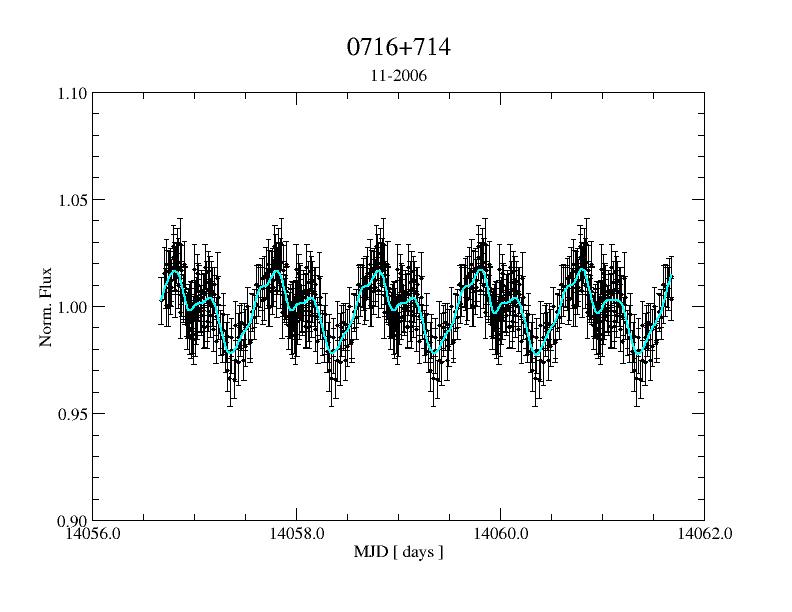}
\hspace{1cm}
   \includegraphics[width=6.5cm]{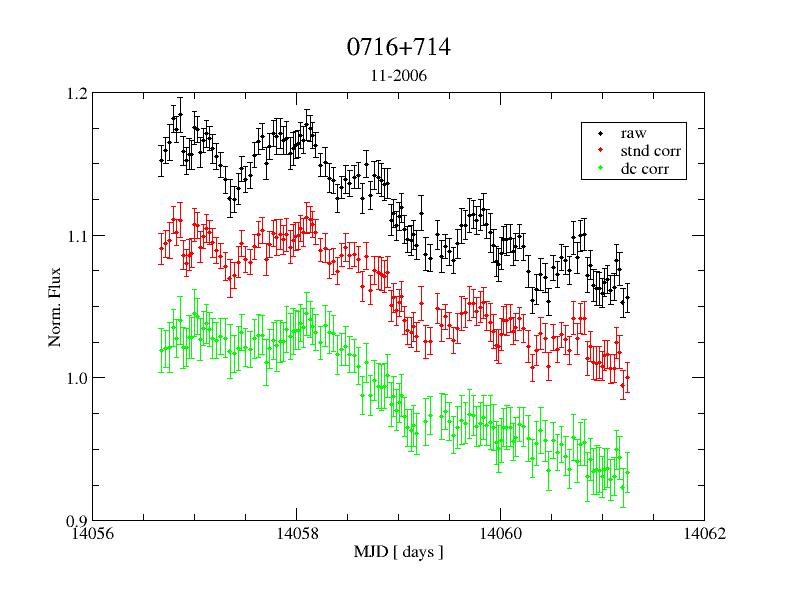}
      \caption{Different steps of the data reduction procedure which was developed for removing spurious periodic modulation. Upper left panel: the variability curve of S5~0716+714, observed in November 2006, after a long-term trend subtraction; upper right panel: the superposition of data obtained on different days of observation; lower left panel: the "daily average modulation" extended to the total duration of the experiment; lower right panel: the variability curves before reduction (black), after the standard procedure for data reduction (red), and after the new procedure (green). For better visualization, the 3 data trains are offset by a small factor.}
         \label{0716_dc}
   \includegraphics[width=8cm]{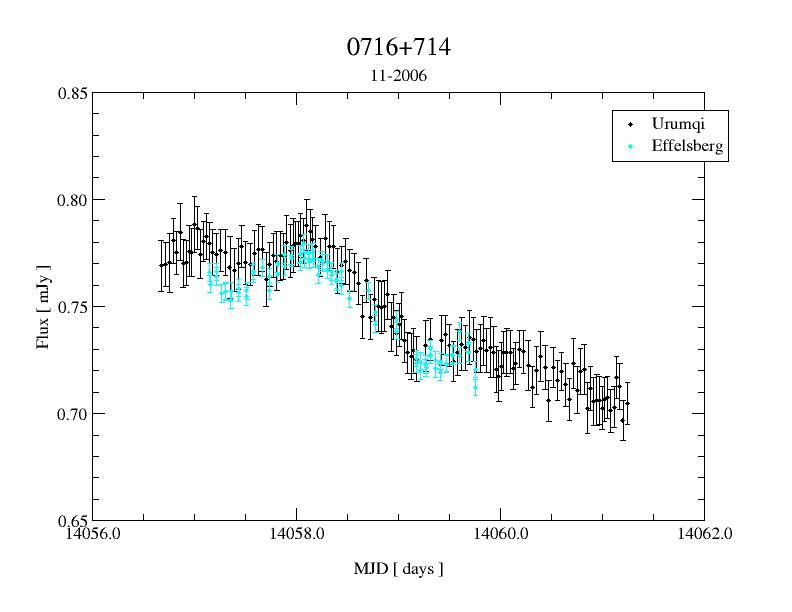}
      \caption{5GHz variability curves of S5~0716+714 from simultaneous observations at Effelsberg and Urumqi in November 2006: the Urumqi curve, after application of the new data reduction procedure, agrees very well with the Effelsberg data.
              }
         \label{0716_dc1}
   \end{figure}

The efficiency of this standard procedure can be evaluated by measuring the residual scatter in the data of the calibrators, in terms of $\frac{\Delta S}{S}$, where $\Delta S$ is the standard deviation. For most of the observations performed in Urumqi, the residual scatter is $0.5-0.7\%$, about 2-3 times worse than in an average Effelsberg experiment - an increase in the noise which is fairly acceptable, considering the difference in the diameter of the antennas. However, a few epochs turned out to be affected by the presence of a systematic $\sim$24-hour modulation in the variability curves, which the standard correction procedures are unable to remove. The nature of the problem is still not completely clear - most likely, it is due to a superposition of two different effects: an inaccurate pointing model\footnote{The problem has now been fixed with replacing the previous pointing model with a new one, including more parameters.}, and a strong day-night effect, causing significant changes in the antenna gain with time. 

In order to remove this spurious variability, a new technique for data reduction has been developed and applied to the affected data. It exploits the fact that the daily modulation is a regular periodic signal. After the removal of the long-term trend, each variability curve is divided in segments of 23.93 hours (i.e., in sideral days), which are then shifted one top of each other. A cubic spline curve is fitted to the data, providing an estimate of the spurious quasi-periodic modulation, which then is expanded to the duration of the whole experiment and is finally used to correct the original variability curve (see Fig. \ref{0716_dc}).\footnote{The division of the curves into sideral days leads to a better overlap than a 24-hour division, showing that the pointing effect is dominating over the day-night effect.}

This technique does not replace the standard procedure described before: it only helps to get rid of systematic daily variations - i.e., it corrects for gain-elevation effect, along with regular pointing displacements and day-night effects. Non-systematic modulations, such as opacity and gain-time effect, have to be removed independently. 

Considering the results of several tests, the technique seems to work very well (see Fig. \ref{0716_dc1}, where the comparison with the Effelsberg data demonstrates the reliability of the method). It requires though observations lasting not less than 3 days, in order to ensure a reasonably good determination of the daily modulation pattern, and a good sampling. These conditions make it not suitable for the reduction of all the datasets.

\section{Results}

The variability time scales of the monitored IDV sources have been determined by means of Structure Function (SF, see \cite{Simonetti}), Discrete Auto-Correlation Function (DACF, \cite{Edelson}) and Mexican-Hat Wavelets, taking the average of the variability time scales resulting from the three methods as the best estimation. Possible correlations between time scale, modulation index\footnote{The modulation index is a parameter used for the estimation of the variability amplitude, defined by $m=100\cdot\frac{\sigma}{<S>}$, where $\sigma$ is the standard deviation of the measured flux density and $<S>$ is time average of the flux density itself.} and average flux have been investigated, but no relevant correlation has been found (see Table \ref{tab_0716}).

 In the sections below, we present preliminary results obtained for two of the main targets of our monitoring campaign, S5~0716+714 and S4~0954+658, and discuss the compatibility between the obtained time scales and the expectations from an isotropic annual modulation model. 

\subsection{S5~0716+714}

S5~0716+714 is a BL~Lac object showing strong variability on time scales from hours to years. It has been the target of several observation campaigns at different wavelengths, and it is so far the only source for which correlated optical-radio intraday variability has been observed. This argument, as already mentioned, would suggest a source-intrinsic origin of IDV. 

In Fig. \ref{Results}, left panel, we plotted the variability time scales versus day of the year, along with two different parameterizations of the isotropic annual modulation model (\cite{Gabanyi}). Unexpectedly, there seem to be some hint of a prolongation of the time scales around day 300. At present, we don't have enough data for establishing whether the slowing down is systematic or just an episode. Further observations are planned for the end of 2007 - hopefully, they will help us to strengthen or discard the hypothesis of annual modulation in S5~0716+714.

In the same plot, in green dots, we reported also the time scale estimations obtained after removing the fast variability (time scales $<0.5d$) from the variability curves of the object. It is interesting to note how small are the differences with respect to the original curves, implying that {\it fast variability plays a marginal role for the determination of the time scales}, and, maybe more important, {\it residual noise in the data, because of its short time coherence, should not seriously affect the results}.
\begin{table}[ht]
\centering
%\vspace{3cm}
\begin{tabular}{|l|c|c|c|c|c|l|c|c|c|c|}
\hline
                  &        & \multicolumn{3}{l}{S5~0716+714} & & & \multicolumn{3}{l}{S4~0954+658} &\\
\hline
 Epoch            & D.o.Y. & Time scale & err & $ M.I. $ & $<Flux>$ & &  Time scale & err & $ M.I. $ & $<Flux>$ \\
                  &       &     [d]    & [d] &   [\%]   & [Jy] & &        [d]     & [d] &  [\%]    & [Jy] \\
\hline
15.03-18.03.2006     &   76  &    0.8     & 0.1 &   1.94   & 0.64 & & 0.9     & 0.1 &   1.60   & 0.92 \\
27.04-01.05.2006     &  120  &    1.3     & 0.2 &   1.43   & 0.64 & & 0.4     & 0.1 &   1.22   & 1.15 \\
09.06-12.06.2006     &  162  &    1.6     & 0.1 &   4.93   & 0.74 & & 0.5     & 0.1 &   1.99   & 1.11 \\
14.07-18.07.2006     &  198  &    0.5     & 0.1 &   2.85   & 0.75 & & 1.5     & 0.2 &   0.90   & 1.02 \\
19.08-25.08.2006     &  235  &    2.0     & 0.3 &   4.48   & 0.85 & & 1.4     & 0.2 &   1.19   & 1.09  \\
23.09-28.09.2006     &  269  &  $>4.0$    & 0.0 &   1.83   & 0.82 & & 1.4     & 0.1 &   1.26   & 1.21\\
17.11-22.11.2006     &  324  &    3.0     & 0.1 &   3.63   & 0.74 & & 2.3     & 0.1 &   1.16   & 1.06\\
18.12-21.12.2006     &  354  &    1.1     & 0.1 &   1.76   & 0.70 & & 0.6     & 0.1 &   0.76   & 0.95\\
25.01-27.01.2007     &   26  &  $>1.6$    & 0.0 &   2.03   & 0.79 & & 0.3     & 0.2 &   0.91   & 1.08 \\
12.02-16.02.2007     &   45  &    2.2     & 0.2 &   2.27   & 0.75 & & 1.8     & 0.3 &   2.28   & 1.06\\
24.03-27.03.2007     &   85  &    1.6     & 0.2 &   2.16   & 0.74 & & 0.9     & 0.1 &   1.27   & 1.29 \\
20.04-24.04.2007     &  113  &    2.0     & 0.1 &   4.35   & 0.74 & & 1.4     & 0.3 &   1.95   & 1.30\\
\hline
\end{tabular}
\caption{Variability characteristics for S5~0716+714 and S4~0954+658, obtained between March 2006 and April 2007. In column 1 and 2 we reported the starting-ending date of the observations and the relative day of the year; in column 3 and 4 (7 and 8, for S4~0954+658), the estimations of the variability time scale and relative errors; in column 5 (9), the modulation index; in column 6 (10), the average flux density.}
\label{tab_0716}
\end{table}

\subsection{S4~0954+658}

The blazar S4~0954+658 is a classical type II-IDV source. Several studies have already been conducted on the source, in both radio and optical bands (see, e.g., \cite{Cimo}, \cite{Papadakis}, \cite{Fuhrmann}), which do not conclusively solve the problem of the origin of its variability, but seem to testify the presence of a scattering screen in front of it.

\begin{figure}
   \centering
   \includegraphics[width=7.5cm]{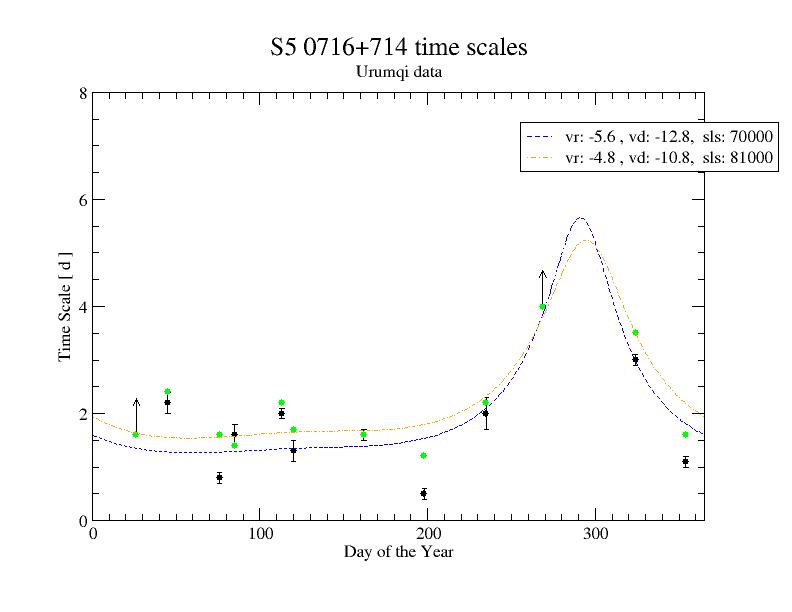}
%\hspace{1cm}
   \includegraphics[width=7.5cm]{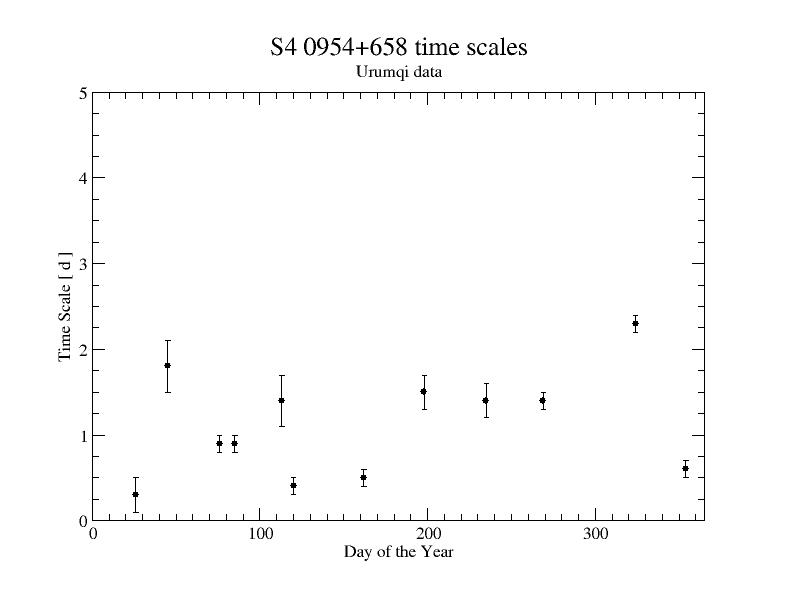}

      \caption{The variability time scales of S5~0716+714 and S4~0954+658 obtained from 12 epochs of Urumqi observations, plotted versus day of the year. In S5~0716+714, a possible slow down of the variability time scales was observed around D.o.Y 300, compatible with and an annual modulation effect. The time scales of S4~0954+658, instead, do not show any systematic change during the monitoring period.}
         \label{Results}
   \end{figure}

The variability time scales inferred from the observations are plotted in Fig. \ref{Results}, right panel. There seems to be no sign of systematic variations of the variability time scales over the year - every attempt to fit the data by means of isotropic or anisotropic (\cite{Gabanyi}) annual modulation models looks doubtful. This result does not rule out ISS as a possible explanation of the IDV observed in S4~0954+658, but the lack of an annual cycle in a strongly active source is a fact which deserves deeper investigation.

\section{Conclusions}

We performed 12 observations at the Urumqi telescope, between March 2006 and April 2007; three epochs of simultaneous Urumqi and Effelsberg observations have revealed a flattering agreement between the variability curves obtained with the two antennas, despite the large difference in diameter, demonstrating that the Chinese telescope is well suited for IDV experiments. 

The datasets have been reduced using both the standard procedure for Effelsberg data reduction and a new method for the removal of a spurious periodic modulation which was due to pointing problems and day-night effects. In order to estimate the characteristic variability time scales, the final data have been analyzed by means of standard techniques of time series analysis, such as Structure Function, Discrete AutoCorrelation Function and Wavelets. For two IDV sources, S5~0716+714 and S4~0954+658, the estimated time scales have been compared with the expectations from an isotropic annual modulation models, leading to the conclusion that hints of annual modulation are present in the variability of S5~0716+714 and not in S4~0954+658.

\section{Acknowledgments}
This work is based on observations with the 100meter telescope
of the MPIfR at Effelsberg (Germany) and with the 25meter
Urumqi telescope of the Urumqi Observatory, National Astronomical
Observatories of the Chinese Academy of Sciences. N. M. and K. \'E. G. 
have been partly supported for this research through a stipend from
the International Max Planck Research School (IMPRS) for Radio and
Infrared Astronomy at the Universities of Bonn and Cologne. X. Liu and H. G. Song are supported by the National Natural Science Foundation (NNSF) of China (10773019)

\end{document}